\def\del{\partial}

\def\abs#1{{\left|{#1}\right|}}
\def\vev#1{\langle #1 \rangle}
\def\del{\partial}

\def\abs#1{{\left|{#1}\right|}}
\def\vev#1{\langle #1 \rangle}

\def\del{\partial}
\def\dslash{\del\kern-0.55em\raise 0.14ex\hbox{/}}

\def\rough#1{\raise.3ex\hbox{$#1$\kern-.75em\lower1ex\hbox{$\sim$}}}

\newcommand{\PRD}[3]{{\it Phys. Rev.} {\bf D{#1}}, {#2} (19{#3})}

\newcommand{\PRDM}[3]{{\it Phys. Rev.} {\bf D{#1}}, {#2} (20{#3})}
\newcommand{\PRL}[3]{{\it Phys. Rev. Lett.} {\bf {#1}}, {#2} (19{#3})}
\newcommand{\PRLM}[3]{{\it Phys. Rev. Lett.} {\bf {#1}}, {#2} (20{#3})}
\newcommand{\NPB}[3]{{\it Nucl. Phys.} {\bf B{#1}}, {#2} (19{#3})}
\newcommand{\NPBM}[3]{{\it Nucl. Phys.} {\bf B{#1}}, {#2} (20{#3})}
\newcommand{\PLB}[3]{Phys. Lett. {\bf B{#1}}, {#2} (19{#3})}
\newcommand{\PLBM}[3]{{\it Phys. Lett.} {\bf B{#1}}, {#2} (20{#3})}

\newcommand{\PTP}[3]{{\it Prog. Theor. Phys.} {\bf {#1}}, {#2} (19{#3})}
\newcommand{\PTPM}[3]{{\it Prog. Theor. Phys.} {\bf {#1}}, {#2} (20{#3})}
\newcommand{\ANN}[3]{{\it Ann. Phys. (N.Y.)} {\bf {#1}}, {#2} (19{#3})}

\newcommand{\MPLM}[3]{{\it Mod. Phys. Lett.} {\bf A{#1}} {#2} (20{#3})}

\newcommand{\jhep}[3]{{\it JHEP} {\bf {#1}} (20{#2}) {#3}}

\renewcommand{\Re}{\mathop{\text{Re}}\nolimits}
\renewcommand{\Im}{\mathop{\text{Im}}\nolimits}
\newcommand{\Li}{\mathop{\text{Li}}\nolimits}
\newcommand{\Res}{\mathop{\text{Res}}}
\newcommand{\tr}{{\text{tr}}}
\documentclass[12pt]{article}
\usepackage{graphicx,amsmath,amssymb,fancybox,amscd}
\usepackage{bm}
\begin{document}
\baselineskip=18pt
\begin{titlepage}
\begin{flushright}
KOBE-TH-11-02\\
OU-HET 703/2011
\end{flushright}
\vspace{1cm}
\begin{center}{\Large\bf 
Gauge-Higgs Unification\par
\vspace{3mm}
in Lifshitz Type Gauge Theory
}
\end{center}
\vspace{0.5cm}
\begin{center}
Hisaki Hatanaka$^{(a)}$
\footnote{E-mail: hatanaka@het.phys.sci.osaka-u.ac.jp},
Makoto Sakamoto$^{(b)}$
\footnote{E-mail: dragon@kobe-u.ac.jp} and
Kazunori Takenaga$^{(c)}$
\footnote{E-mail: takenaga@kumamoto-hsu.ac.jp}
\end{center}
\vspace{0.2cm}
\begin{center}
${}^{(a)}$ {\it Department of Physics,
Osaka University,
Toyonaka, Osaka 560-0043, Japan}
\\[0.2cm]
${}^{(b)}$ {\it Department of Physics, Kobe University, 
Rokkodai, Nada, Kobe 657-8501, Japan}
\\[0.2cm]
${}^{(c)}$ {\it Faculty of Health Science,
Kumamoto Health Science University,
Izumi-machi, Kumamoto 861-5598, Japan}
\end{center}
\vspace{1cm}
\begin{abstract}
We discuss the gauge-Higgs unification in a framework of Lifshitz
type gauge theory.
We study a higher dimensional gauge theory on $R^{D-1}\times S^{1}$
in which the normal second (first) order derivative terms for  
scalar (fermion) fields in the action are replaced by higher order
derivative ones for the direction of the extra dimension.
We provide some mathematical tools to evaluate a one-loop
effective potential for the zero mode of the extra component 
of a higher dimensional gauge field and clarify how the higher
order derivative terms affect the standard form of the effective
potential.
Our results show that they can make the Higgs mass heavier and
change its vacuum expectation value drastically.
Some extensions of our framework are briefly discussed.
\end{abstract}
\end{titlepage}
%
%
%
%
%
\newpage
\section{Introduction}
%
%
%
Gauge theories in higher dimensions are a promising candidate
beyond the Standard Model.
Such theories turn out to possess unexpectedly rich properties
that shed new light and give a deep understanding on high
energy physics. In fact, it has been shown that new mechanisms
of gauge symmetry breaking \cite{manton, fairlie, sherk,
hosotani, higgsless}, spontaneous supersymmetry 
breaking \cite{susy},
and breaking of translational invariance \cite{translation1,
translation2} can occur, and that various phase structures arise
in field theoretical models
on certain topological manifolds \cite{higgshosotani1, higgshosotani2,
highT}. 
Furthermore, new diverse scenarios of solving the hierarchy 
problem have been proposed in \cite{rs, manybrane, 
nagasawasakamoto, sakamototakenaga1}.
\par
 
Since the origin of gauge symmetry breaking is still an unsolved
problem, it is worth pursuing an alternative mechanism to
mimic the Standard Model Higgs.
In this paper, we focus on the gauge-Higgs unification that
the \lq\lq Higgs\rq\rq\ field arises from an extra dimensional
component of a higher dimensional gauge field \cite{manton, fairlie, 
hosotani}.
Higher dimensional gauge invariance forbids any Higgs mass term
at tree level and the Higgs potential can be generated through
quantum corrections.
In the gauge-Higgs unification, the Higgs field corresponds to
the Wilson line phase and the
effective potential never suffers from ultraviolet (UV) divergences
due to the nonlocal property of the Wilson line phase.
As a result, the vacuum expectation value and the mass of the 
Higgs field derived from the effective potential are finite
and calculable.
This is a very attractive feature of the gauge-Higgs unification.
A flat compactification on a circle $S^{1}$, however, leads
to the light Higgs mass problem in normal settings of 
gauge-Higgs unification models \cite{lightHiggsmass}.
A compactification on a warped extra dimension may solve the
problem \cite{lightHiggsmass_warped}.
In this paper, we take an alternative approach to solve it.
\par
 
In the standard gauge-Higgs unification, the kinetic terms of
bosonic (fermionic) fields in the action are taken to be the
second (first) order derivatives for extra dimensions as well as
the Minkowski space-time.
Lorentz invariance does not, however, require the order of 
the derivatives to be the same for the extra dimensions and the 
Minkowski space-time because it is explicitly broken by the
compactification.
We may thus have an opportunity to introduce higher
derivative terms for the direction of extra dimensions.
\par
 
Recently, Ho{\v r}ava \cite{Horava} has proposed 
an interesting idea to make gravity theory power-counting 
renormalizable in 4-dimensions.
His idea is to treat space and time non-relativistically and 
the anisotropy between them is given by the introduction of
higher spatial derivative terms characterized by the 
dynamical critical exponent $z$.
A number of studies have been made on the Ho{\v r}ava-Lifshitz
gravity.\footnote{
Studies at an early stage have been given in Ref.\cite{cosmology}
for cosmological applications, Ref.\cite{BH} for black hole
physics and Ref.\cite{theoreticalaspect} for theoretical aspects.
}
Field Theories with higher spatial derivative terms have also
been investigated on various subjects of renormalization 
\cite{HLrenormalization}, non-gauge models \cite{HLnongauge},
QED \cite{HLQED, Casimir}, Yang-Mills theory \cite{HLYM},
SUSY \cite{HLSUSY, lightvelocity} and the Standard Model
extension \cite{HLSM}.
All Ho{\v r}ava-Lifshitz type models lose Lorentz invariance
at high energies but it is expected that it would emerge
at low energies as an accidental symmetry \cite{Horava}.
The recovery of Lorentz invariance at low energies is, however,
a nontrivial problem since in a theoretical point of view
there is no reason that different particles possess the same
limiting speed (i.e. the common light velocity $c$) in the absence
of Lorentz invariance.\footnote{
A resolution to this problem has been proposed 
in Ref.\cite{lightvelocity}.
}
Problems of Ho{\v r}ava-Lifshitz type theory have been reviewed in
Ref.\cite{HLproblem}.
\par

In this paper, we introduce higher derivative terms
only for the direction of extra dimensions and keep Lorentz
invariance intact for the Minkowski space-time.
Thus, our models have the anisotropy between the Minkowski
space-time and the extra dimensions but not between space
and time.
This situation will be rather close to the original idea 
proposed by Lifshitz \cite{Lifshitz}.
Higher order derivative terms would become important if 
coefficients of lower order derivative terms happen to vanish.
This is indeed the case when the system lies at a Lifshitz
point \cite{Lifshitzpoint}.
We thus consider the gauge-Higgs unification in a higher
dimensional gauge theory at a Lifshitz point, 
though we will not investigate whether our models would lie at a 
Lifshitz point, but simply assume it in this paper.
\par
 
As noted before, the effective potential for the Higgs field,
which is a zero mode of an extra dimensional component of a
higher dimensional gauge field, is finite and free from UV
divergence.
This UV insensitivity does not, however, imply that higher
order derivative terms are irrelevant to the effective potential.
It turns out that they bring about quantitative changes of the
effective potential and play an important role to solve the
light Higgs problem.
To show this is one of the main purposes of this paper.
We will also develop some mathematical tools to compute 
one-loop effective potentials.
This is another purpose of this paper.
\par
 
This paper is organized as follows:
In the next section, we explain our setup of the gauge-Higgs
unification in a framework of Lifshitz type gauge theory.
In section 3, we evaluate the one-loop effective potential
for the zero mode of the extra dimensional component of the
gauge field and discuss how higher derivative terms affect
the effective potential.
In section 4, we present a five dimensional $SU(2)$ model to
demonstrate properties found in section 3, explicitly.
In section 5, we extend the results in section 3 to the Lifshitz
type gauge theory on $R^{D-2}\times S^{1}\times S^{1}$.
Section 6 is devoted to conclusions and discussions.
Technical details will be found in appendices.

%
%
%
%
%
\section{Lifshitz Type Gauge Theory on 
$\bm{R^{D-1}\!\times\! S^{1}}$}
%
%
%
In this section, we present our setup of the gauge-Higgs
unification in a framework of Lifshitz type gauge theory.
To this end, we consider a $D$-dimensional $SU(N)$ gauge
invariant theory compactified on a circle.\footnote{
Since the gauge fixing and the ghost terms are irrelevant
in our discussions, we have omitted them in Eq.(\ref{action}).
}
%
\begin{align}
\label{action}
S &= \int d^{D-1}x \int_{0}^{L} dy\, \Bigl\{
     -\frac{1}{4}F_{\mu\nu}F^{\mu\nu} - \frac{1}{2}F_{\mu y}F^{\mu y}
     + \overline{\Psi} \bigl( i\gamma^{\mu}D_{\mu}
       + l^{z-1}(i\gamma^{y}D_{y})^{z} + M_{F}\bigr)\Psi\notag\\
  &\hspace{40mm}  
     + \Phi^{*} \bigl( D^{\mu}D_{\mu}
       -l^{2(z-1)}(-D_{y}^{\ 2})^{z} - M_{S}^{\ 2}\bigr)\Phi
       - V(\Phi) \Bigr\}\,, 
\end{align}
%
where $x^{\mu}\ (\mu=0,1,\cdots, D-2)$
denotes the $(D-1)$-dimensional Minkowski space-time coordinate and
$y$ is the coordinate of the extra dimension on the circle $S^{1}$ of 
the circumference $L$.
The covariant derivatives $D_{\mu}$ and $D_{y}$ for the field
$\Psi$ belonging to the representation $\cal{R}$ of $SU(N)$ are
given by
%
\begin{align}
\label{covariantderivative_mu}
D_{\mu}\Psi 
 &= (\partial_{\mu} + igA_{\mu}^{c}T^{c}_{(\cal{R})})\Psi\,,\\
\label{covariantderivative_y} 
D_{y}\Psi 
 &= (\partial_{y} + igA_{y}^{c}T^{c}_{(\cal{R})})\Psi\,,
\end{align}
%
where $T^{c}_{(\cal{R})}\ (c=1,2,\cdots,N^{2}-1)$ is a generator
of $SU(N)$ in the representation $\cal{R}$.
\par
 
We should here make several comments on the higher derivative
terms in the action (\ref{action}).
We have introduced the higher derivative terms only for the 
direction of the extra dimension and assumed that there are no
higher derivative terms such as 
$(\partial_{\mu}\partial^{\mu})^{m} (\partial_{y})^{n}$.
This implies that there is no possibility to add any higher
order derivative terms for the gauge field in a gauge invariant
way because $F_{yy}=0$ and the term $(F_{\mu y}F^{\mu y})^{z}$
for $z>1$
would produce higher derivatives with respect to the 
extra dimensional coordinate and the Minkowski space-time ones as well.
We could add lower order derivatives of the extra dimensional
one such as $(i\gamma^{y}D_{y})^{n}\ (n=1,2,\cdots,z-1)$
for the fermion and $(D_{y})^{2n}$ for the scalar, but we have 
omitted those terms because we would like to clarify the effects
of the higher derivative terms.
It could be justified if the system sits on a Lifshitz point,
where the lower derivative terms become irrelevant.
We will not, however, pursue such a possibility.
We simply keep only the highest order derivative terms 
labeled by the dynamical critical exponent $z$ 
and investigate the theory at a Lifshitz point.
It should be noticed that a length parameter $l$ has been introduced
to adjust the dimension of the higher derivative terms.
The ratio $l/L$ turns out to be very important in determining the
vacuum expectation value and the mass of the Higgs field.
\par
 
Since the extra dimension is compactified on the circle,
we  have to specify boundary conditions on the fields.
We here take the periodic boundary condition for all the fields, i.e.\footnote{
Although the periodicity of the fields on a circle $S^{1}$
may require the boundary conditions (4) \textit{up to
gauge transformations}, we here restrict a class of
the gauge parameter $\Lambda(x,y)$ to the periodic
function $\Lambda(x,L)=\Lambda(x,0)$, so that all the
fields are assumed to be periodic with respect to the
coordinate $y$ on $S^{1}$.
}
%
\begin{align}
%
A_{\mu}(x,L) &= A_{\mu}(x,0)\,,\notag\\
A_{y}(x,L) &= A_{y}(x,0)\,,\notag\\
\Psi(x,L) &= \Psi(x,0)\,,\notag\\
\label{bc_Phi}
\Phi(x,L) &= \Phi(x,0)\,.
\end{align}
%
The extension to other boundary conditions will be straightforward.
\par
 
In the gauge-Higgs unification, the zero mode of $A_{y}$ plays a
role of the Higgs field and the vacuum expectation value
$\vev{A_{y}}$ can be determined dynamically as the minimum of the
effective potential, which is induced through radiative corrections.
The purpose of the next section is to compute the one-loop
effective potential for the zero mode of $A_{y}$ from the action
(\ref{action}) with the boundary conditions (\ref{bc_Phi}).

%
%
%
%
%
\section{One-Loop Effective Potential}
%
%
%
In this section, we evaluate the one-loop effective potential 
for the dynamical variable $\vev{A_y}$ from the action (\ref{action})
of the 5d Lifshitz type gauge theory on $R^{D-1}\times S^{1}$.
%
\begin{align}
\label{effectivepotential}
V^{R^{D-1}\times S^{1}_{z}}_{\textrm{eff}}(\alpha)
 = V^{R^{D-1}\times S^{1}_{z}}_{\textrm{scalar}}(\alpha; M_{S})
  + V^{R^{D-1}\times S^{1}_{z}}_{\textrm{fermion}}(\alpha; M_{F})
  + V^{R^{D-1}\times S^{1}_{z}}_{\textrm{gauge}}(\alpha)\,,
\end{align}
%
where the subscript $z$ of $S^{1}$ indicates the Lifshitz type
higher derivatives of the dynamical critical exponent
$z$ for the $S^{1}$ direction.
$V^{R^{D-1}\times S^{1}_{z}}_{\textrm{scalar}}, 
V^{R^{D-1}\times S^{1}_{z}}_{\textrm{fermion}}$
and $V^{R^{D-1}\times S^{1}_{z}}_{\textrm{gauge}}$ denote 
the contributions from scalar, fermion and gauge loop diagrams, 
respectively and the dimensionless variable $\alpha$ is defined by
%
\begin{align}
\label{a}
\alpha \equiv g \vev{A_{y}} \frac{L}{2\pi}\,.
\end{align}
%
In the following, we first evaluate the effective potential for the 
massless matter and then for the massive one.
%
%
%
%
%
\subsection{Massless Matter}
%
%
%
In this subsection, we consider the massless scalar and fermion 
fields with $M_{S}=M_{F}=0$.
Let us start with 
$V^{R^{D-1}\times S^{1}_{z}}_{\textrm{scalar}}(\alpha; M_{S}=0)$
which comes from the scalar loop diagram
%
\begin{align}
\label{V_s}
V^{R^{D-1}\times S^{1}_{z}}_{\textrm{scalar}}(\alpha; M_{S}=0)
 = \int \frac{d^{D-1}p_{E}}{(2\pi)^{D-1}}\ 
    \frac{1}{L} \sum^{\infty}_{m=-\infty}
    \textrm{tr}_{(\cal{R})}\biggl[
    \ln\Bigl( p_{E}^{\ 2} + l^{2(z-1)}\Bigl(
    \frac{2\pi(m+\alpha)}{L}\Bigr)^{2z}\Bigr)\biggr]\,,
\end{align}
%
where $p_{E}$ denotes the $(D-1)$-dimensional Euclidean momentum.
When the Lifshitz scalar $\Phi$ which propagates the loop belongs
to the representation $\cal{R}$ of the gauge group, the trace
in Eq.(\ref{V_s}) should be taken over the gauge indices with
respect to the representation $\cal{R}$ and $\alpha$ should be
expressed as $\alpha=\alpha^{c}T^{c}_{(\cal{R})}$.
We note that $V^{R^{D-1}\times S^{1}_{z}}_{\textrm{scalar}}$
is an unrenormalized quantity, so that it should be renormalized
to obtain a finite expression.
\par
 
For $z=1$, which corresponds to the normal kinetic term, the
techniques have already been well established to obtain the
finite expression from Eq.(\ref{V_s}).
Some of the techniques for $z=1$, however, turn out not to apply
for the case of $z>1$.
Thus, we need to develop
mathematical tools to compute the one-loop effective
potential for $z>1$.
This is one of the purposes of the present paper.
\par
 
Let us now evaluate $V^{R^{D-1}\times S^{1}_{z}}_{\textrm{scalar}}$.
We first rewrite Eq.(\ref{V_s}) into the form
%
\begin{align}
\label{V_s_01}
V^{R^{D-1}\times S^{1}_{z}}_{\textrm{scalar}}(\alpha; M_{S}=0)
 &= -\Bigl(\frac{1}{4\pi}\Bigr)^{\frac{D-1}{2}}
    \frac{1}{L} \sum^{\infty}_{m=-\infty}
    \int^{\infty}_{0} dt\,t^{-\frac{D+1}{2}}\notag\\
 &\qquad\qquad   \times\textrm{tr}_{(\cal{R})}\biggl[
    \exp\Bigl\{ -t\, l^{2(z-1)}\Bigl(
    \frac{2\pi(m+\alpha)}{L}\Bigr)^{2z}\Bigr\}\biggr]\,,
\end{align}
%
where we have used the standard formulas
%
\begin{align}
\label{formula_01}
&\ln A 
 = -\frac{d}{ds}\Bigl( \frac{1}{\Gamma(s)}\int^{\infty}_{0} dt\,
     t^{s-1} e^{-At}\Bigr)\Bigr\vert_{s\to +0}\,,\\
\label{formula_02}
&\frac{d}{ds}\Bigl( \frac{t^{s}}{\Gamma(s)}\Bigr)\Bigr\vert_{s\to +0}
 = 1\,,\\
\label{formula_03}
&\int^{\infty}_{-\infty}\frac{dp}{2\pi} e^{-tp^{2}}
 = \sqrt{\frac{1}{4\pi t}}\,.
\end{align}
%
For $z=1$, we can use the Poisson summation formula
%
\begin{align}
\label{Poissonformula}
\sum^{\infty}_{m=-\infty} 
   \exp\Bigl\{-t\Bigl(\frac{2\pi(m+\alpha)}{L}\Bigr)^{2}\Bigr\}
 = \frac{L}{\sqrt{4\pi t}} \sum^{\infty}_{n=-\infty}
   \exp\Bigl\{-\Bigl(\frac{nL}{4t}\Bigr)^{2}-2\pi i n\alpha\Bigr\}
\end{align}
%
and then arrive at the well known renormalized expression\footnote{
The UV divergent term of $n=0$ has been removed from the summation
in Eq.(\ref{V_s_02}) to renormalize the one-loop effective potential.
}
%
\begin{align}
\label{V_s_02}
V^{R^{D-1}\times S^{1}_{z=1}}_{\textrm{scalar}}(\alpha; M_{S}=0)
 &= -2\frac{\Gamma(D/2)}{\pi^{D/2}}\frac{1}{L^{D}}
    \sum^{\infty}_{n=1} \textrm{tr}_{(\cal{R})}\Bigl[
    \frac{\cos(2\pi n\alpha)}{n^{D}}\Bigr]\,.
\end{align}
%
\par
 
For $z>1$, we may use the original Poisson summation formula, i.e.
%
\begin{align}
\label{originalPoissonformula}
\sum^{\infty}_{m=-\infty} f(m+\alpha)
 = \sum^{\infty}_{n=-\infty} e^{2\pi in\alpha}
   \int^{\infty}_{-\infty}du f(u) e^{-2\pi i nu}\,,
\end{align}
%
in place of Eq.(\ref{Poissonformula}).
With the help of the above formula and the relation 
$\Gamma(s) = \int^{\infty}_{0} dt\, t^{s-1}e^{-t}$,
we can have a renormalized expression
%
\begin{align}
\label{V_s_03}
V^{R^{D-1}\times S^{1}_{z}}_{\textrm{scalar}}(\alpha; M_{S}=0)
 &= -\Bigl(\frac{1}{4\pi}\Bigr)^{\frac{D-1}{2}}
    \Gamma(-(D-1)/2)\, l^{(z-1)(D-1)}\frac{1}{L}
    \Bigl(\frac{2\pi}{L}\Bigr)^{z(D-1)}
    \notag\\
 &\qquad\times\sum^{\infty}_{n=-\infty}{}^{\!\!\!\!\prime}\,
    \textrm{tr}_{(\cal{R})}
    \bigl[\exp\{2\pi in\alpha\}\bigr] \int^{\infty}_{-\infty}du
    \vert u\vert^{z(D-1)} e^{-2\pi inu}\,,
\end{align}
%
where the prime of the summation denotes that the $n=0$ 
contribution, which corresponds to the UV divergent part,
should be removed from the summation to get a finite result
of the effective potential.
Furthermore, using the formula of the Fourier transform 
\cite{IwanamiII}
%
\begin{align}
\label{Fouriertransform}
\int^{\infty}_{-\infty} du \vert u\vert^{\nu} e^{-2\pi inu}
 = -2\, \sin\Bigl(\frac{\pi \nu}{2}\Bigr) \,
    \frac{\Gamma(1+\nu)}{\vert2\pi n\vert^{1+\nu}}
\end{align}
%
for $\nu\ne \cdots,-3,-1,0,2,4,\cdots$, we obtain 
%
\begin{align}
\label{V_s_04}
V^{R^{D-1}\times S^{1}_{z}}_{\textrm{scalar}}(\alpha; M_{S}=0)
 &= -8\Bigl(\frac{1}{4\pi}\Bigr)^{\frac{D+1}{2}}
    \Gamma(-(D-1)/2) \Gamma(1+z(D-1)) \sin \Bigl(\frac{\pi z(D-1)}{2}\Bigr)
    \notag\\
 &\qquad \times\frac{l^{(z-1)(D-1)}}{L^{1+z(D-1)}}
    \sum^{\infty}_{n=1}{}\,
    \textrm{tr}_{(\cal{R})}
    \Bigl[\frac{\cos (2\pi n\alpha)}{n^{D+(z-1)(D-1)}}\Bigr]\,.
\end{align}
%
It should be noticed that the expression (\ref{V_s_04}) would be
ill defined for $D=$ odd or $z=$ even because the formula 
(\ref{Fouriertransform}) cannot apply for those cases and also
the factor $\Gamma(-(D-1)/2)$ diverges for $D=$ odd.
To avoid these problems, we regard the space-time dimension $D$
as a real (or complex) number and define the expression 
(\ref{V_s_04}) by the analytic continuation of $D$.
It then turns out that the final result is finite and well defined
for any positive integers $D$ and $z$:
%
\begin{align}
\label{V_s_05}
V^{R^{D-1}\times S^{1}_{z}}_{\textrm{scalar}}(\alpha; M_{S}=0)
 &= -2\Bigl(\frac{1}{\pi}\Bigr)^{\frac{D}{2}}
    \Gamma(D/2) \frac{\Gamma(D+(z-1)(D-1))}{\Gamma(D)}
    h_{D,z} \notag\\
 &\qquad \times \frac{1}{L^{D}}
    \Bigl(\frac{l}{L}\Bigr)^{(z-1)(D-1)}
    \sum^{\infty}_{n=1}
    \textrm{tr}_{(\cal{R})}
    \Bigl[\frac{\cos (2\pi n\alpha)}{n^{D+(z-1)(D-1)}}\Bigr] \,,
\end{align}
%
where
%
\begin{align}
\label{h}
h_{D,z}
  \equiv \lim_{s\to D} 
    \frac{\sin \bigl(\pi z(s-1)/2\bigr)}
         {\sin\bigl(\pi(s-1)/2\bigr)}
 = \left\{
   \begin{array}{cl}
    z & \textrm{for}\ D=\textrm{odd},\ z=\textrm{odd},\\
    z(-1)^{(D-1)/2} & \textrm{for}\ D=\textrm{odd},\ z=\textrm{even},\\
    (-1)^{(z-1)/2} & \textrm{for}\ D=\textrm{even},\ z=\textrm{odd},\\
    0 & \textrm{for}\ D=\textrm{even},\ z=\textrm{even}.
   \end{array}\right.
\end{align}
%
Here, we have used the formulas of the gamma function
%
\begin{align}
\label{formula04}
&\Gamma(s)\,\Gamma(1-s)
 =\frac{\pi}{\sin(\pi s)}\,,\\
\label{formula05}
&\Gamma(2s)
 =\frac{2^{2s}}{2\sqrt{\pi}}\,\Gamma(s)\,\Gamma(s+1/2)\,.
\end{align}
%
For $z=1$, Eq.(\ref{V_s_05}) is found to exactly reproduce the 
expression (\ref{V_s_02}).
This will give a consistency check of our result (\ref{V_s_05}).
\par
 
To confirm the validity of the result (\ref{V_s_05}) furthermore,
we would like to derive Eq.(\ref{V_s_05}) in a different way.
To this end, we rewrite Eq. (\ref{V_s_01})  as
\begin{eqnarray}
\lefteqn{
V^{R^{D-1} \times S_z^1}_{\rm scalar} (\alpha; M_S = 0)
} \nonumber\\
&=&
- \frac{\Gamma(- \tfrac{D-1}{2})}{(4\pi)^{\frac{D-1}{2}}}\frac{l^{(z-1)(D-1)} }{L}
\left( \frac{2\pi}{L} \right)^{z(D-1)} 
\tr_{({\cal R})}
\sum_{m=-\infty}^{\infty}\left(
{1\over \abs{m+\alpha}}\right)^{-z(D-1)}.
\label{effpot-zeta}
\end{eqnarray}
Then, using the Hurwitz zeta function
%
\begin{align}
\label{zetafunction}
\zeta(s,\alpha)
  = \sum^{\infty}_{m=0} \frac{1}{(m+\alpha)^{s}}
    \qquad \textrm{for}\quad\ 0<\alpha\le 1\,,
\end{align}
we have
\begin{eqnarray}
\lefteqn{
V^{R^{D-1} \times S_z^1}_{\rm scalar} (\alpha; M_S = 0)} 
\nonumber\\
&=&
- \frac{\Gamma(- \tfrac{D-1}{2})}{(4\pi)^{\frac{D-1}{2}}}\frac{l^{(z-1)(D-1)} }{L}
\left( \frac{2\pi}{L} \right)^{z(D-1)} 
\nonumber\\&&
\times \tr_{({\cal R})}
\{ \zeta(-z(D-1),\alpha) + \zeta(-z(D-1),1-\alpha) \}.
\label{effpot-zeta1}
\end{eqnarray}
The formula \cite{Gradshteyn}
%
\begin{align}
\label{Hurwitzformula}
\zeta(s,\alpha)
  = \frac{2\Gamma(1-s)}{(2\pi)^{1-s}}
     \Bigl\{ \sin \Bigl(\frac{\pi s}{2}\Bigr)
             \sum^{\infty}_{n=1}\frac{\cos(2\pi n\alpha)}{n^{1-s}}
           + \cos \Bigl(\frac{\pi s}{2}\Bigr)
             \sum^{\infty}_{n=1}\frac{\sin(2\pi n\alpha)}{n^{1-s}}
     \Bigr\}\,,
\end{align}
%
for ${\textrm {Re}}~s<0$, leads to the same result (\ref{V_s_05}),
as it should be.
It is interesting to note that UV divergence has been removed
automatically from the effective potential thanks to the zeta
function regularization.
We will further find other different derivations of Eq.(\ref{V_s_05})
in Appendix A.
\par
 
We have succeeded to compute the scalar contribution to the
one-loop effective potential for the zero mode of $A_{y}$.
It is now easy to obtain the fermionic contribution 
$V^{R^{D-1}\times S^{1}_{z}}_{\textrm{fermion}}(\alpha; M_{F}=0)$.
The result is 
%
\begin{align}
\label{V_s_06}
V^{R^{D-1}\times S^{1}_{z}}_{\textrm{fermion}}(\alpha; M_{F}=0)
 &= +2^{[D/2]}\Bigl(\frac{1}{\pi}\Bigr)^{\frac{D}{2}}
    \Gamma(D/2) \frac{\Gamma(D+(z-1)(D-1))}{\Gamma(D)}
    h_{D,z} \notag\\
 &\qquad \times \frac{1}{L^{D}}
    \Bigl(\frac{l}{L}\Bigr)^{(z-1)(D-1)}
    \sum^{\infty}_{n=1}
    \textrm{tr}_{(\cal{R})}
    \Bigl[\frac{\cos (2\pi n\alpha)}{n^{D+(z-1)(D-1)}}\Bigr] \,.
\end{align}
%
The difference between the scalar and fermion contributions appears
in the overall factors.
The factor $-2$ in $V^{R^{D-1}\times S^{1}_{z}}_{\textrm{scalar}}$
is replaced by $+2^{[D/2]}$ in
$V^{R^{D-1}\times S^{1}_{z}}_{\textrm{fermion}}$.
The difference of the overall sign comes from that of the spin
and statistics for bosons and fermions.
The factor $2$ in $V^{R^{D-1}\times S^{1}_{z}}_{\textrm{scalar}}$
and $2^{[D/2]}$ in $V^{R^{D-1}\times S^{1}_{z}}_{\textrm{fermion}}$
are just the number of (on-shell) degrees of freedom for a complex
scalar and a  Dirac spinor, respectively.\footnote{
For a real scalar or a Weyl/Majorana spinor, the factor should be
replaced by $1$ or $2^{[D/2]-1}$.
}
\par
 
The gauge field contribution to the one-loop effective potential in our
model is nothing but that of $z=1$, i.e.
%
\begin{align}
\label{V_s_07}
V^{R^{D-1}\times S^{1}_{z}}_{\textrm{gauge}}(\alpha)
 &= -(D-2)\frac{\Gamma(D/2)}{\pi^{D/2}}\frac{1}{L^{D}}
    \sum^{\infty}_{n=1} \textrm{tr}_{(\textrm{adj})}\Bigl[
    \frac{\cos(2\pi n\alpha)}{n^{D}}\Bigr]\,.
\end{align}
%
This result will lead to an interesting observation 
to determine the vacuum expectation value of $\alpha$.
Since $V^{R^{D-1}\times S^{1}_{z}}_{\textrm{gauge}}$ is independent
of $l$, the gauge loop contribution will become less important than
the scalar and fermion loop ones when the ratio $l/L$ is large.
On the other hand, it becomes important when $l/L$ is small enough.
\par
 
Before closing this subsection, we would like to make comments on
the $z$-dependence of $V^{R^{D-1}\times S^{1}_{z}}_{\textrm{scalar}}$
and $V^{R^{D-1}\times S^{1}_{z}}_{\textrm{fermion}}$.
For $z=1$, the infinite summation over $n$ is given by
the form $\sum_{n}\cos(2\pi n\alpha)/n^{D}$.
The power of $n$ is just the total space-time dimension $D$.
For general $z$, the power of $n$ is replaced by $D+(z-1)(D-1)$.
This can be understood from an anisotropic scaling point of view.
It follows from the action (\ref{action}) that we may introduce the
anisotropic scaling for the space-time coordinates as \cite{Horava}
%
\begin{align}
\label{anisotropicscaling}
x^{\mu}\ 
  &\to\ b^{z}\,x^{\mu},\quad (\mu=0,1,2,\cdots,D-2)\,,
  \notag\\
y\ &\to\ b\,y\,.
\end{align}
%
Then, the effective dimension $\overline{D}$ of the system may
be defined from the scaling of the volume element as
%
\begin{align}
\label{effectiveD}
[d^{D-1}x\, dy]\ \to\ b^{\overline{D}}\, [d^{D-1}x\, dy]
\end{align}
%
with $\overline{D}=z(D-1)+1=D+(z-1)(D-1)$, which agrees with the
power of $n$ in Eq.(\ref{V_s_05}) and Eq.(\ref{V_s_06}).
\par

Another interesting observation is the overall sign of 
$V^{R^{D-1}\times S^{1}_{z}}_{\textrm{scalar}}$
and $V^{R^{D-1}\times S^{1}_{z}}_{\textrm{fermion}}$.
For $z=1$, the overall sign is determined by the spin and statistics
but not the space-time dimension $D$.
In fact, $h_{D,z}=1$ for $z=1$ (and any $D$) but it can change the
sign for $z>1$ without changing the spin and statistics.
Furthermore, the contribution to the one-loop effective potential vanishes
for $z=$ even and $D=$ even.\footnote{
A similar property has been found in the Casimir effect \cite{Casimir}.
}
This is not the case for $z=1$.
\par

The other important observation is that a huge numerical factor
can arise for $z>1$.
The numerical values of $\Gamma(D+(z-1)(D-1))/\Gamma(D)$ for
$D=5$ and $z=1,2,3,4$ are, for instance, listed below:
%
\begin{align}
\label{hugefactor}
\frac{\Gamma(5+4(z-1))}{\Gamma(5)}
 \sim \left\{
   \begin{array}{cl}
    1 & \textrm{for}\ z=1,\\
    1.7\times10^3 & \textrm{for}\ z=2,\\
    2.9 \times 10^7 & \textrm{for}\ z=3,\\
    8.7 \times 10^{11} & \textrm{for}\ z=4.
   \end{array}\right.
\end{align}
%
Another source of a large/small number will come from the factor
$(l/L)^{(z-1)(D-1)}$, which depends on the ratio of $l$ and $L$.
In the gauge-Higgs unification, the Higgs mass squared will be given by
%
\begin{align}
\label{Higgsmass}
m_{H}^{\ 2}
 \sim \frac{\partial^{2} V_{\textrm{eff}}}{\partial A_{y}^{\ 2}}
 \Bigr\vert_{A_{y}=\vev{A_{y}}}\,.
\end{align}
%
Those numerical factors can change the value of the Higgs mass
drastically.\footnote{It has also been reported in \cite{Panico-Serone-Wulzer} that the Higgs 
mass can be heavy due to the effect of the violation of the five-dimensional Lorentz invariance.} 
Therefore, the overall factor of $V_{\textrm{eff}}$ is very sensitive to the magnitude of $m_{H}$.
The Higgs mass prediction in Lifshitz type gauge theory is expected to
be quite different from ordinary gauge-Higgs unification models with $z=1$.
%
%
%
%
%
\subsection{Massive Matter}
In this subsection, we evaluate the one-loop effective potential 
for massive matter.
The computations of 
$V^{R^{D-1}\times S^{1}_{z}}_{\textrm{scalar}}(\alpha; M_{S})$
and $V^{R^{D-1}\times S^{1}_{z}}_{\textrm{fermion}}(\alpha; M_{F})$
will involve a number of technical manipulations as compared to the
massless case.
We here give only the result of 
$V^{R^{D-1}\times S^{1}_{z}}_{\textrm{scalar}}(\alpha; M_S)$ 
for $D=$ odd:
%
\begin{align}
\label{V_s_08}
&V^{R^{D-1}\times S^{1}_{z}}_{\textrm{scalar}}(\alpha; M_S)\notag\\
 &\quad = -
    \frac{4z}{(4\pi)^{(D-1)/2}\Gamma(\frac{D-1}{2})}\ \frac{1}{L}
    \sum^{\frac{D-3}{2}}_{q=0} {}_{\frac{D-3}{2}}C_{q}\,
    (-M_S^{2})^{\frac{D-3}{2}-q}
    \Bigl(\frac{l^{z-1}}{L^{z}}\Bigr)^{2q+2} 
    \Gamma\bigl(z(2q+2)\bigr)
    \notag\\
 &\qquad   \times
    {\rm tr}_{({\cal R})}\sum^{z}_{j=1} \Re
    \Bigl[ (-i\omega_{j})^{z(2q+2)}    
    \sum^{z(2q+2)-1}_{k=0} \frac{(\rho_{j})^{k}}{k!}
           \Li_{z(2q+2)-k+1}(e^{-\rho_{j}+i2\pi \alpha})\Bigr]\,, 
\end{align}
%
where 
%
\begin{align}
\label{omega}
\omega_{j}
 &= \exp\Bigl(i\frac{2j-1}{2z}\,\pi\Bigr)\,,\\
\label{rho}
\rho_{j}
 &= \frac{i}{\omega_{j}}\frac{L M_S^{1/z}}{l^{(z-1)/z}}\,,\\
\label{Li}
\Li_{s}(x)
 &= \sum^{\infty}_{n=1}\frac{x^{n}}{n^{s}}\,.
\end{align}
%
The details will be found in Appendix B.
For $D=5$ and $z=1$, Eq.~\eqref{V_s_08} reads
\begin{eqnarray}
V_{\rm scalar}^{R^4 \times S_{z=1}^1}(\alpha; M_S)
&=&
- \frac{1}{2\pi^2 L^5} \tr_{({\cal R})} \Re \bigl[
 3 \Li_{5} (e^{-LM_S + i 2 \pi \alpha})
\nonumber\\ && \quad 
 +3 L M_S \Li_{4} (e^{-LM_S + i 2 \pi \alpha})
+  L^2 M_S^2 \Li_{3} (e^{-LM_S + i 2 \pi \alpha})
\bigr],
\end{eqnarray}
which is consistent with the result in Ref.~\cite{Delgado}.
It is not difficult to verify that Eq.(\ref{V_s_08}) 
reduces to Eq.(\ref{V_s_05}) in the limit of $M_S\to 0$
with the relation $\sum_{j=1}^{z}(-i\omega_{j})^{z(D-1)} = h_{D,z}$, 
as it should be.
The fermionic contribution 
$V^{R^{D-1}\times S^{1}_{z}}_{\textrm{fermion}}(\alpha; M_F)$ can be
obtained by replacing the factor $-2$ in 
$V^{R^{D-1}\times S^{1}_{z}}_{\textrm{scalar}}(\alpha; M_S)$
by $+2^{[D/2]}$.
It follows from the expression (\ref{V_s_08}) that 
the contribution from a particle with
the bulk mass $M_S$ will be suppressed exponentially if
%
\begin{align}
\label{masssuppression}
M_S >\!\!> \Bigl( \frac{l}{L}\Bigr)^{z-1}\frac{1}{L}\,.
\end{align}
%
This observation imply that Lifshitz particles of $z>1$ even
with $M_S \gtrsim 1/L$ may contribute to the effective potential
if $l/L>\!\!>1$.
This should be compared with the case of $z=1$.
Only massive particles with $M_S \lesssim 1/L$ contribute to
the effective potential. 
%
%
%
\section{An $\bm{SU(2)}$ Gauge Model with an Adjoint Fermion}
%
%
%
Let us consider an $SU(2)$ gauge theory on $R^{4}\times S_{z}^{1}$ coupled 
to a massless adjoint fermion whose kinetic term has the Ho{\v r}ava type higher 
derivative. We are interested in the gauge symmetry breaking patterns 
and the mass of the adjoint scalar field which is originally 
the component gauge field for the compactified direction.  
\par

The one-loop effective potential is useful tool 
in order to study them. 
In the present case, from the discussions in the previous section, 
the effective potential is given by
\begin{align}
V_{\textrm{eff}}^{R^{4}\times S^{1}_{z}}(\alpha;M=0)
&={1\over \pi^{5\over 2}}
{\Gamma\left({5\over 2}\right)\over L^5}\biggl[
-3\sum_{n=1}^{\infty}{2\over n^5}\biggl(1+\cos(2\pi n(2\alpha))\biggr)
4h_{D=5, z}
\nonumber\\
& \quad+\left({l\over L}\right)^{4(z-1)}{\Gamma(4z+1)\over \Gamma(5)}
\sum_{n=1}^{\infty}{2\over n^{4z+1}}\biggl(1+\cos(2\pi n(2\alpha))\biggr)
\biggr].
\label{shiki1}
\end{align}
The first line in Eq. (\ref{shiki1}) is the contribution from the
gauge (and ghost) fields and the second one is the one from the
adjoint fermion. The Wilson line phase $\alpha$ is related by the vacuum expectation 
value $\vev{A_y}$,
\begin{equation}
gL\vev{A_y}=2\pi~{\rm diag.}(\alpha, -\alpha)
\label{shiki2}
\end{equation}
and the factor $h_{D=5, z}$ is given, from (\ref{h}), by 
\begin{equation}
h_{D=5, z}=z.
%
\label{shiki3}
\end{equation}
The $L$ is the length of the circumference of the circle $S^1$.
We see that the magnitude of the higher derivative, which is 
shown by the ratio $l/L$, plays the role of changing the size 
of the contribution from the adjoint fermion to 
the effective potential.  The gauge symmetry breaking patterns 
through the Wilson line phases (Hosotani mechanism) has 
been studied extensively \cite{tsuika}, and we understand 
that the pattern depends on the matter content of the theory.
\par
If the scale $l$ is set to zero, then, the one-loop effective potential 
has the contribution from the gauge (and the ghost) fields
alone. It has been known that the minimum of the potential is located at 
\begin{equation}
\alpha=0\qquad \left(\mbox{mod}~~{1\over 2}\right),
\label{shiki4}
\end{equation}   
for which the $SU(2)$ gauge symmetry is not broken. On the other
hand, if we take the scale $l$ to be 
large enough, then, the effective potential
is dominated by the adjoint fermion. In this case, the minimum of the 
effective potential is
\begin{equation}
\alpha=0.25,
\label{shiki5}
\end{equation}
so that the $SU(2)$ gauge symmetry is broken to $U(1)$.
\par

The above observation suggests that the gauge symmetry breaking 
patterns depend on the ratio $l/L$. 
%
%
%
\begin{figure}[ht]
\begin{center}
\includegraphics[width=9cm]{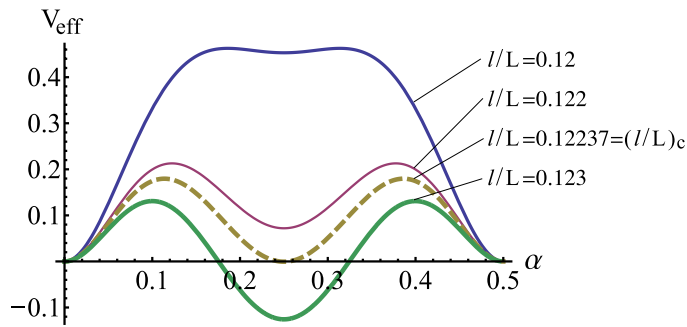}
\caption{The behavior of the one-loop effective potential with respect to
the ratio ${l\over L}=0.12, 0.122, 0.12237, 0.123$ with
$z=2$. According to the change of the ratio, the vacuum 
configuration changes from $\alpha=0$ (mod $1/2$) 
to $\alpha=0.25$. The degenerate vacuum appears at 
the critical ratio $({l\over L})_c=0.12237$, above which the vacuum configuration 
is $\alpha=0.25$, where gauge symmetry is broken down to $U(1)$.}
\end{center}
\end{figure}
%
%
%
Actually, the vacuum structure changes 
according to the magnitude of $l/L$, as shown in Fig.$1$. We find that for $z=2$ the 
degenerate vacuum appear at the critical ratio,
\begin{equation}
\left({l\over L}\right)_c=0.12237.
\label{shiki6}
\end{equation}
For larger value than the critical ratio, the $SU(2)$ gauge symmetry is broken to
$U(1)$, while the smaller value than the critical value, the $SU(2)$ gauge
symmetry is unbroken. The behavior of the vacuum expectation
value (VEV) $\alpha$ with respect to $l/L$ is depicted in
Fig.$2$, where we observe that the VEV jumps at the critical 
ratio, hence the phase transition is first order. 
\begin{figure}[ht]
\begin{center}
\includegraphics[width=9cm]{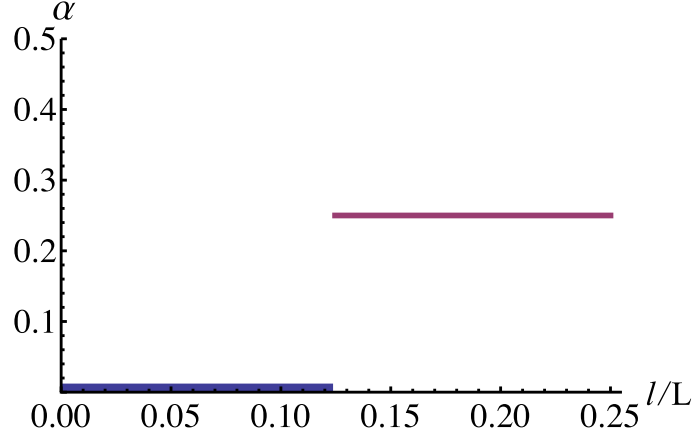}
\caption{The behavior of the vacuum expectation value (VEV) $\alpha$ with 
respect to ${l\over L}$. At the critical ratio $({l\over L})_c=0.12237$, the VEV 
jumps, and the phase transition is first order.}
\end{center}
\end{figure}
\par

Let us next discuss the mass of the adjoint scalar field which is originally 
the component gauge field for the $S^1$ direction. The mass of the adjoint 
Higgs scalar is obtained by the second derivative 
of the effective potential evaluated at the vacuum,\footnote{
Let us note that the adjoint Higgs scalar can be massive even 
if the gauge symmetry is not broken.}
\begin{eqnarray}
m_{H}^2&=&{\del^2V_{\textrm{eff}}^{R^{4}\times S^{1}_{z}}
(\alpha)\over \del \vev{A_y^3}^2}\bigg|_{\rm vac}
=\left({gL\over 4\pi} \right)^2~
{\del^2V_{\textrm{eff}}^{R^{4}\times S^{1}_{z}}
(\alpha)\over \del\alpha^2}\bigg|_{\rm vac}
\nonumber\\
&=&{3~{\bar g}^2\over 2\pi^2 L^2}
\biggl[
3\sum_{n=1}^{\infty}{1\over n^3}\cos(2\pi n(2\alpha))
\nonumber\\
&&-4z{\Gamma(4z+1)\over \Gamma(5)}
\biggl({l\over L} \biggr)^{4(z-1)}~\sum_{n=1}^{\infty}
{1\over n^{4z-1}}\cos(2\pi n(2\alpha))
\biggr]\bigg|_{\rm vac}.
\label{shiki7}
\end{eqnarray}
where we have defined the 4d gauge coupling constant
${\bar g}\equiv g/\sqrt{L}$. Hereafter we take $z=2$ for numerical studies.
\par

Let us study the case of $\alpha=0.25$, for which the gauge symmetry 
is broken to $U(1)$. The adjoint Higgs mass is given by
\begin{equation}
m_{H}^2={3~{\bar g}^2\over 2\pi^2 L^2} 
\biggl[
-{9\over 4}~\zeta(3)+
4\times 2\times 
{8!\over 4!}
\left({l\over L}\right)^4
\times {63\over 64}~\zeta(7)
\biggr],
\label{shiki8}
\end{equation}
where we have used the Riemann's zeta function,
\begin{equation}
\sum_{n=1}^{\infty}{(-1)^n\over n^{4k-1}}=-(1-2^{2-4k})\zeta(4k-1)
\label{shiki9}
\end{equation}
with $k=1, 2$. The mass squared $m_{H}^2$ is positive definite for the range
of $(l/L)_c < l/L$, where the vacuum configuration is given 
by $\alpha=0.25$. We note that the Gamma 
function $\Gamma(4z+1)$ is sizable enhancement for the mass.
The adjoint scalar mass depends on the ratio $l/L$, which can also
enhance the mass of the adjoint scalar field.
\par
The ratio between the adjoint Higgs scalar and the lightest massive gauge
boson is given by
\begin{equation}
{m_{H}\over M_W^{(0)}}
={\bar g}\sqrt{{3\over 2\pi^4}
\left(-{9\over 4}\zeta(3)
+13230\times \zeta(7)\times \biggl({l\over L}\biggr)^4\right)},
\label{shiki10}
\end{equation}
where we have used the expression for the 
the massive Kaluza-Klein gauge bosons after the gauge symmetry 
breaking,
\begin{equation}
M_W^{(n)2}
=\left(2\pi\over L\right)^2
\left(n+2\alpha\right)^2\Bigg|_{\alpha=0.25}\quad \mbox{for}\quad n=0,1,2,\cdots.
\label{shiki11}
\end{equation}
We obtain the mass of the adjoint Higgs scalar field
for various values of $l/L$ \footnote{If we take 
$l/L\sim0.393~(0.4)$ and ${\bar g}\sim 0.65$, we have 
$m_{H}\sim 115~(119)$ GeV. }
\begin{equation}
{m_{H}\over M_W^{(0)}}
={\bar g}\times 
\left\{
\begin{array}{cl}
0.54 & \mbox{for}~~{l\over L}=0.2,\\[0.2cm]
2.28 & \mbox{for}~~{l\over L}=0.4,\\[0.2cm]
9.17 & \mbox{for}~~{l\over L}=0.8,\\[0.2cm]
14.3 & \mbox{for}~~{l\over L}=1.0,\\[0.2cm]
%
%
20.6 & \mbox{for}~~{l\over L}=1.2,\\[0.2cm]
%
%
\end{array}\right.
\label{shiki12}
\end{equation}
The adjoint Higgs mass is generated through loop effects and is natural to 
be light compared with the massive gauge boson appeared after the 
gauge symmetry breaking. In the present case, however, thanks to 
the arbitrary scale $l$ and the Gamma function $\Gamma(4z+1)$, the
adjoint scalar mass can be heavier than the massive 
gauge boson, as shown above.   
\par
The Higgs mass is also affected by the dynamical critical
exponent $z$. In order to see it, let us study the 
ratio between the adjoint Higgs scalar and the lightest 
massive gauge boson with respect to $z$ for fixed value of $l/L$.
We take $l/L=0.25$ as an illustration, for 
which the $SU(2)$ gauge symmetry is broken to $U(1)$. We obtain, from 
(\ref{shiki7}) and (\ref{shiki11}), that
\begin{equation}
{m_{H}\over M_W^{(0)}}
={\bar g}\times 
\left\{
\begin{array}{cl}
0.12 & \mbox{for}~~z=1,\\[0.2cm]
0.87 & \mbox{for}~~z=2,\\[0.2cm]
7.50 & \mbox{for}~~z=3,\\[0.2cm]
113.15 & \mbox{for}~~z=4.\\[0.2cm]
%
%
%
%
\end{array}\right.
\label{shiki13}
\end{equation}
We observe that the Higgs mass highly depends on $z$ and is enhanced by the
effect of the higher derivative, as pointed out in the previous section. 
\section{Higher Dimensional Extension}
%
%
%
In this section, we extend the previous analysis in section 3
to the Lifshitz type gauge theory on 
$R^{D-2}\times S^{1}_{z_{1}}\times S^{1}_{z_{2}}$, and show that
the effective potential on 
$R^{D-2}\times S^{1}_{z_{1}}\times S^{1}_{z_{2}}$
is given by the sum of an effective potential on 
$R^{D-2}\times R^{1}_{z_{1}}\times S^{1}_{z_{2}}$
and infinitely many effective potentials for Kaluza-Klein modes
on $R^{D-2}\times S^{1}_{z_{1}}$.
We here focus on a massless scalar contribution to the one-loop
effective potential
%
\begin{align}
\label{V_s_09}
&V^{R^{D-2}\times S^{1}_{z_{1}}\times S^{1}_{z_{2}}}_{\textrm{scalar}}
   (\alpha_{1},\alpha_{2}; M_S=0)\notag\\
&\qquad = -\Bigl(\frac{1}{4\pi}\Bigr)^{(D-2)/2}
    \Bigl(\frac{1}{L_{1}} \sum^{\infty}_{m_{1}=-\infty}\Bigr)
    \Bigl(\frac{1}{L_{2}} \sum^{\infty}_{m_{2}=-\infty}\Bigr)    
    \int^{\infty}_{0} dt\,t^{-D/2}\notag\\
&\qquad\quad   \times\textrm{tr}_{(\cal{R})}\biggl[
    \exp\Bigl\{ -t \Bigl[l_{1}^{2(z_{1}-1)}\Bigl(
    \frac{2\pi(m_{1}+\alpha_{1})}{L_{1}}\Bigr)^{2z_{1}}
    +l_{2}^{2(z_{2}-1)}\Bigl(
     \frac{2\pi(m_{2}+\alpha_{2})}{L_{2}}\Bigr)^{2z_{2}}
     \Bigr]\Bigr\}\biggr]\,,
\end{align}
%
where the scalar action has been assumed to include 
$l_{1}^{2(z_{1}-1)}(D_{y_{1}})^{2z_{1}}$ and
$l_{2}^{2(z_{2}-1)}(D_{y_{2}})^{2z_{2}}$ for $S^{1}_{z_{1}}$
and $S^{1}_{z_{2}}$ directions, respectively.
The $L_{1}$ and $L_{2}$ are the circumferences of $S^{1}_{z_{1}}$
and $S^{1}_{z_{2}}$, and $\alpha_{1}$ and $\alpha_{2}$ are defined by
$\alpha_{1}=g\vev{A_{y_{1}}}L_{1}/(2\pi)$ and 
$\alpha_{2}=g\vev{A_{y_{2}}}L_{2}/(2\pi)$.
We note that $\alpha_{1}$ and $\alpha_{2}$ are assumed to commute each other
to minimize the tree level potential
$g^{2}\textrm{tr}_{(\textrm{adj})} [A_{y_{1}}, A_{y_{2}}]^{2}$.
Using the Poisson summation formula (\ref{originalPoissonformula}),
we obtain the UV finite expression
%
\begin{align}
\label{V_s_10}
&V^{R^{D-2}\times S^{1}_{z_{1}}\times S^{1}_{z_{2}}}_{\textrm{scalar}}
   (\alpha_{1},\alpha_{2}; M_S=0)\notag\\
 &\qquad= -\Bigl(\frac{1}{4\pi}\Bigr)^{\frac{D-2}{2}}
    \frac{1}{L_{1}L_{2}}
    \sum^{\infty}_{n_{1},n_{2}=-\infty}{}^{\hspace{-4mm}\prime}\hspace{2mm}
    \int_{0}^{\infty}dt\,t^{-D/2}
    \int_{-\infty}^{\infty}du_{1} \int_{-\infty}^{\infty}du_{2} \notag\\
 &\qquad\quad
   \times\textrm{tr}_{(\cal{R})}\biggl[
    \exp\Bigl\{ -t \Bigl[l_{1}^{2(z_{1}-1)}\Bigl(
    \frac{2\pi u_{1}}{L_{1}}\Bigr)^{2z_{1}}
    +l_{2}^{2(z_{2}-1)}\Bigl(
     \frac{2\pi u_{2}}{L_{2}}\Bigr)^{2z_{2}}\Bigr]\notag\\
 &\qquad\quad
     -2\pi i\bigl(n_{1}u_{1}+n_{2}u_{2}\bigr)
     +2\pi i\bigl(n_{1}\alpha_{1}+n_{2}\alpha_{2}\bigr)
     \Bigr\}\biggr]\,,
\end{align}
%
where the prime of the summation denotes that the contribution of
$n_{1}=n_{2}=0$ has to be removed from the summation to subtract
the UV divergent part from the effective potential.
The summation over $n_{1}$ and $n_{2}$ may be rearranged as follows:
%
\begin{align}
\label{summation}
\sum_{n_{1}, n_{2}=-\infty}^{\infty}{}^{\hspace{-4.5mm}\prime}
 &\equiv \sum_{n_{1}=-\infty}^{\infty}
         \sum_{n_{2}=-\infty}^{\infty}
         \bigl( 1- \delta_{n_{1},0}\delta_{n_{2},0}\bigr)\notag\\
 &= \sum_{n_{1}=-\infty}^{\infty}
         \sum_{n_{2}=-\infty}^{\infty}
         \bigl\{ \delta_{n_{1},0}\bigl( 1- \delta_{n_{2},0}\bigr)
                + \bigl( 1-\delta_{n_{1},0}\bigr)\bigr\}\notag\\
 &= \sum_{n_{1}=-\infty}^{\infty}\delta_{n_{1},0}
     \sum_{n_{2}=-\infty}^{\infty}{}^{\hspace{-2mm}\prime}  
    + \sum_{n_{1}=-\infty}^{\infty}{}^{\hspace{-2mm}\prime}\ 
       \sum_{n_{2}=-\infty}^{\infty}\,.           
\end{align}
%
Thus, we can rewrite Eq.(\ref{V_s_10}) as
%
\begin{align}
\label{V_s_11}
&V^{R^{D-2}\times S^{1}_{z_{1}}\times S^{1}_{z_{2}}}_{\textrm{scalar}}
   (\alpha_{1},\alpha_{2}; M=0)\notag\\
 &\qquad= -\Bigl(\frac{1}{4\pi}\Bigr)^{\frac{D-2}{2}}
    \frac{1}{L_{1}L_{2}}
    \sum^{\infty}_{n_{2}=-\infty}{}^{\!\!\!\!\prime}  
    \int_{0}^{\infty}dt\,t^{-D/2}
    \int_{-\infty}^{\infty}du_{1} \int_{-\infty}^{\infty}du_{2} \notag\\
 &\qquad\qquad
   \times\textrm{tr}_{(\cal{R})}\biggl[
    \exp\Bigl\{ -t \Bigl[l_{1}^{2(z_{1}-1)}\Bigl(
    \frac{2\pi u_{1}}{L_{1}}\Bigr)^{2z_{1}}
    +l_{2}^{2(z_{2}-1)}\Bigl(
     \frac{2\pi u_{2}}{L_{2}}\Bigr)^{2z_{2}}\Bigr]\notag\\
 &\qquad\qquad
     -2\pi i n_{2}u_{2} + 2\pi i n_{2}\alpha_{2}
     \Bigr\}\biggr]\notag\\
 &\qquad -\Bigl(\frac{1}{4\pi}\Bigr)^{\frac{D-2}{2}}
    \frac{1}{L_{1}L_{2}}
    \sum^{\infty}_{n_{1}=-\infty}{}^{\!\!\!\!\prime}  \ 
    \sum^{\infty}_{n_{2}=-\infty} 
    \int_{0}^{\infty}dt\,t^{-D/2}
    \int_{-\infty}^{\infty}du_{1} \int_{-\infty}^{\infty}du_{2} \notag\\
 &\qquad\qquad
   \times\textrm{tr}_{(\cal{R})}\biggl[
    \exp\Bigl\{ -t \Bigl[l_{1}^{2(z_{1}-1)}\Bigl(
    \frac{2\pi u_{1}}{L_{1}}\Bigr)^{2z_{1}}
    +l_{2}^{2(z_{2}-1)}\Bigl(
     \frac{2\pi u_{2}}{L_{2}}\Bigr)^{2z_{2}}\Bigr]\notag\\
 &\qquad\qquad
     -2\pi i\bigl(n_{1}u_{1}+n_{2}u_{2}\bigr)
     +2\pi i\bigl(n_{1}\alpha_{1}+n_{2}\alpha_{2}\bigr)
     \Bigr\}\biggr]\,.
\end{align}
%
The first term on the r.h.s. of Eq.(\ref{V_s_11}) can be expressed,
with the change of variable $p_{1}=2\pi u_{1}/L_{1}$, as
%
\begin{align}
\label{1st_term}
&-\Bigl(\frac{1}{4\pi}\Bigr)^{\frac{D-2}{2}}\frac{1}{L_{2}}
    \sum^{\infty}_{n_{2}=-\infty}{}^{\!\!\!\!\prime}  
    \int_{0}^{\infty}dt\,t^{-D/2}
    \int_{-\infty}^{\infty}\frac{dp_{1}}{2\pi}
    \int_{-\infty}^{\infty}du_{2} \notag\\
 &\qquad\quad
   \times\textrm{tr}_{(\cal{R})}\biggl[
    \exp\Bigl\{ -t \Bigl[l_{1}^{2(z_{1}-1)}p_{1}^{2z_{1}}
     +l_{2}^{2(z_{2}-1)}\Bigl(
     \frac{2\pi u_{2}}{L_{2}}\Bigr)^{2z_{2}}\Bigr]
     -2\pi i n_{2}u_{2} + 2\pi i n_{2}\alpha_{2}
     \Bigr\}\biggr]\notag\\
 &\quad\equiv 
  V^{R^{D-2}\times R^{1}_{z_{1}}\times S^{1}_{z_{2}}}_{\textrm{scalar}}
  (\alpha_{2}; M=0)\,.
\end{align}
%
It turns out that 
$V^{R^{D-2}\times R^{1}_{z_{1}}\times S^{1}_{z_{2}}}_{\textrm{scalar}}
(\alpha_{2}; M=0)$ has a clear geometrical meaning.
It is nothing but the one-loop effective potential of a 
$D$-dimensional Lifshitz type gauge theory on 
$R^{D-2}\times R^{1}_{z_{1}}\times S^{1}_{z_{2}}$ (but not
$R^{D-2}\times S^{1}_{z_{1}}\times S^{1}_{z_{2}}$) where
a Lifshitz type higher derivative labeled by $z_{1}$
appears for one of the coordinates in $R^{D-1}$
(i.e. $R^{1}_{z_{1}}$)
as well as for the coordinate of $S^{1}_{z_{2}}$.
\par

The second term on the r.h.s. of Eq.(\ref{V_s_11}) can be
expressed, by using the Poisson summation formula reversely, as
%
\begin{align}
\label{2nd_term}
&\frac{1}{L_{2}}\sum_{m_{2}=-\infty}^{\infty}\frac{-1}{(4\pi)^{(D-2)/2}}
    \frac{1}{L_{1}}
    \sum^{\infty}_{n_{1}=-\infty}{}^{\!\!\!\!\prime}  \ 
    \int_{0}^{\infty}dt\,t^{-D/2}
    \int_{-\infty}^{\infty}du_{1}  \notag\\
 &\qquad\qquad
   \times\textrm{tr}_{(\cal{R})}\biggl[
    \exp\Bigl\{ -t \Bigl[l_{1}^{2(z_{1}-1)}\Bigl(
    \frac{2\pi u_{1}}{L_{1}}\Bigr)^{2z_{1}}
     + M_{m_{2}}^{\ 2} \Bigr]
     -2\pi i n_{1}u_{1}+2\pi i n_{1}\alpha_{1} \Bigr\}\biggr]\notag\\
 &\quad\equiv 
  \frac{1}{L_{2}}\sum_{m_{2}=-\infty}^{\infty}
   V^{R^{D-2}\times S^{1}_{z_{1}}}_{\textrm{scalar}}
  (\alpha_{1}; M_{m_{2}})\,,
\end{align}
%
where 
%
\begin{align}
\label{M_m2}
M_{m_{2}} \equiv
 l_{2}^{\,z_{2}-1}\Bigl\vert
   \frac{2\pi(m_{2}+\alpha_{2})}{L_{2}}\Bigr\vert^{z_{2}}\,.
\end{align}
%
Again, $V^{R^{D-2}\times S^{1}_{z_{1}}}_{\textrm{scalar}}
(\alpha_{1}; M_{m_{2}})$ turns out to have a clear geometrical meaning.
It is nothing but the one-loop effective potential of a
$(D-1)$-dimensional Lifshitz type gauge theory (but not 
$D$-dimensional one) on $R^{D-2}\times S^{1}_{z_{1}}$ where the scalar
action contains the Lifshitz type higher derivative of 
$z_{1}$ for the $S^{1}_{z_{1}}$ direction with the bulk mass
$M_{m_{2}}$ that corresponds to the Kaluza-Klein mass of the
mode $m_{2}$.
Thus, we found that\footnote{
A similar structure has been found in gauge theories on 
$R^{D-1}\times S^{1}$ at finite temperature \cite{highT}.
}
%
\begin{align}
\label{V_s_12}
&V^{R^{D-2}\times S^{1}_{z_{1}}\times S^{1}_{z_{2}}}_{\textrm{scalar}}
   (\alpha_{1},\alpha_{2}; M=0)\notag\\
 &\quad=
  V^{R^{D-2}\times R^{1}_{z_{1}}\times S^{1}_{z_{2}}}_{\textrm{scalar}}
    (\alpha_{2}; M=0)
  + \frac{1}{L_{2}}\sum_{m_{2}=-\infty}^{\infty}
     V^{R^{D-2}\times S^{1}_{z_{1}}}_{\textrm{scalar}}
      (\alpha_{1}; M_{m_{2}})\,.
\end{align}
%
Since the above observation is expected to hold for fermion and
gauge fields, the one-loop effective potential of the Lifshitz
type gauge theory on $R^{D-2}\times S^{1}_{z_{1}}\times S^{1}_{z_{2}}$
turns out be written into a decomposition form similar to
Eq.(\ref{V_s_12}).
We also expect that the effective potential for a Lifshitz type
gauge theory on $R^{D-N}\times S^{1}_{z_{1}}\times\cdots
\times S^{1}_{z_{N}}$ has a similar structure, though
we will not proceed furthermore.
%
%
%
%
%
\section{Conclusions and Discussions}
%
%
%
We have investigated the Lifshitz type gauge theory in the
gauge-Higgs unification.
The Lifshitz scalar and fermion possess the kinetic terms of 
the higher derivatives labeled by $z$ for the direction
of the extra dimension.
We have succeeded to evaluate the one-loop effective potential
for the zero mode of the extra dimensional component of 
the gauge field and found that it heavily depends on the
dynamical critical exponent $z$ of the Lifshitz particles.
The overall sign of the effective potential can change with
respect to $z$ irrespective of the spin and statistics, and
furthermore the one-loop effective potential turns out to 
vanish for even $D$ and $z$.
A huge numerical factor arises for $z>1$ and this property
may solve the light Higgs mass problem.
It has also been found that the length parameter $l$ plays
an important role in determining the vacuum expectation value
of the Higgs field as well as the magnitude of the Higgs mass.
\par

To obtain the finite expression of the one-loop effective
potential, we have extended the space-time dimension $D$ to a  
real (or complex) number and taken it to be the original
integral value only at the final stage.
To confirm this procedure, we have provided the
mathematical tools and derived the one-loop effective potential
in the several different ways.
We have also studied the one-loop effective potentials numerically
in the 5d $SU(2)$ model, and confirmed various peculiar properties of
our models.
\par

The analysis for the Lifshitz type gauge theory on 
$R^{D-1}\times S^{1}_{z}$ has been extended to the higher extra
dimensions $S^{1}_{z_{1}}\times S^{1}_{z_{2}}$.
An interesting observation is that the one-loop effective
potential on $R^{D-2}\times S^{1}_{z_{1}}\times S^{1}_{z_{2}}$
can be expressed by the sum of a one-loop effective potential
on $R^{D-2}\times R^{1}_{z_{1}}\times S^{1}_{z_{2}}$ and
infinitely many one-loop ones on $R^{D-2}\times S^{1}_{z_{1}}$,
which is one-dimension lower than the original space-time
dimension $D$, coming from Lifshitz Kaluza-Klein particles on
$S^{1}_{z_{2}}$.
A similar decomposition property will hold for the Lifshitz
type gauge theory on the higher extra dimension
$S^{1}_{z_{1}}\times S^{1}_{z_{2}}\times \cdots \times S^{1}_{z_{N}}$.
\par
Another extension is to replace the circle $S_z^1$ with the orbifold $S_z^1/Z_2$. We can 
apply the methods developed here to this case. Hence, it is important and interesting 
to study the gauge symmetry breaking and Higgs mass in a context of Lifshitz type
gauge theories on $S_z^1/Z_2$.
\par
In this paper, we have assumed the higher derivative terms to
present only for the direction of the extra dimensions.
Ho{\v r}ava's original idea \cite{Horava} is, however, to demand
the anisotropy between time and space coordinates to make 
gravity theory power-counting renormalizable.
According to the Ho{\v r}ava's spirit, we may replace the second
order derivatives of all the spatial coordinates of $R^{D-1}\times S^{1}$
by higher derivatives to make the theory renormalizable.
The power-counting renormalizability requires the mass dimension of 
the gauge coupling to be non-negative.
This implies that the dynamical critical exponent $z$ should be greater
than or equal to $D-3$.
The one-loop effective potential coming from such a Ho{\v r}ava-Lifshitz
massless scalar loop may be given by
%
\begin{align}
\label{V_s_13}
&V^{R^{1}\times R^{D-2}_{z}\times S^{1}_{z}}_{\textrm{scalar}}
   (\alpha; M=0)\notag\\
&\qquad = -\int\frac{d^{D-2}\bm{p}}{(2\pi)^{D-2}}
           \frac{1}{L} \sum^{\infty}_{m=-\infty}
           \int^{\infty}_{0} dt\,t^{-1}\frac{1}{\sqrt{4\pi t}}\notag\\
&\qquad\qquad   \times\textrm{tr}_{(\cal{R})}\biggl[
    \exp\Bigl\{ -t\, l^{2(z-1)}\Bigl(
     \bigl(\bm{p}^{2}\bigr)^{z} 
     + \Bigl(\frac{2\pi(m+\alpha)}{L}\Bigr)^{2z}
     \Bigr)\Bigr\}\biggr]\notag\\
&\qquad = -2z\Bigl(\frac{1}{\pi}\Bigr)^{1/2}C_{z}
         \frac{\sin\bigl(\pi(z+D-2)/2\bigr)\Gamma\bigl(z+D-2\bigr)}
         {\sin\bigl(\pi(z+D-2)/(2z)\bigr)\Gamma\bigl((z+D-2)/(2z)\bigr)}    
         \notag\\
&\qquad\qquad  \times\frac{1}{L^{D}}\Bigl(\frac{l}{L}\Bigr)^{z-1}
        \sum_{n=1}^{\infty}
        \textrm{tr}_{({\cal R})}\biggl[\frac{\cos(2\pi n\alpha)}{n^{z+D-1}}\biggr]\,,  
\end{align}
%
where 
%
\begin{align}
\label{C}
C_{z} \equiv
  \int\frac{d^{D-2}\bm{p}}{(2\pi)^{D-2}}\ e^{-(\bm{p}^{2})^{z}}.
\end{align}
%
We should emphasize that standard gauge-Higgs unification models
are not renormalizable because the total space-time dimension $D$
is greater than 4.
Thus, only a very limited class of physical quantities such as 
the Higgs mass are finite and calculable, so that the predictability
of the theory is quite restricted.
Since the Ho{\v r}ava-Lifshitz type gauge theory is power-counting
renormalizable, any physical quantities can, in principle, be
computed with finite values at the cost of Lorentz symmetry violation.
It would be of great interest to investigate Ho{\v r}ava-Lifshitz
type gauge theory in more details.
\vspace{5mm}
\begin{center}
{\bf Acknowledgement}
\end{center}
This work is supported in part by a Grant-in-Aid for Scientific Research
(No. 22540281 and No. 20540274 (M.S.), No. 21540285 (K.T.)) from the Japanese 
Ministry of Education, Science, Sports and Culture. 
The authors would like to thank 
Professors M. Kato, N. Maru and H. So for valuable discussions.
\vspace*{1cm}
%
%
%
\vspace{5mm}
\appendix
\section{Other Methods to Evaluate the Effective Potential}
%
In the followings, we show the alternative ways to reproduce the result \eqref{V_s_05}.
\subsection{A method with contour Integrals}
\begin{figure}[htbp]
\centerline{\includegraphics[width=.3\linewidth]{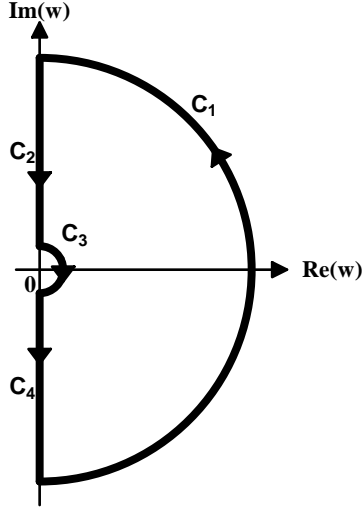}}
\caption{The integration path stretched on the positive-real plane.}\label{contour-two}
\end{figure}

Eq.~\eqref{V_s_05} can be reproduced in a way similar to \cite{Garriga:2000jb}.
First, in terms of contour integrals, Eq.~\eqref{effpot-zeta} can be written as
\begin{eqnarray}
\lefteqn{V_{\rm scalar}^{R^{D-1}\times S_z^1}(\alpha; M_S =0) }
\nonumber\\
&=& -
\frac{\Gamma\left(-{D-1\over 2}\right)}{(4 \pi)^{(D-1)\over 2}}
\frac{l^{(z-1)(D-1)}}{L}
\left(\frac{2\pi}{L}\right)^{z(D-1)}
\frac{1}{2\pi i} 
\textrm{tr}_{(\cal{R})}
\oint_C d w \, w^{z(D-1)}\frac{d F(w,\alpha)/d w}{F(w,\alpha)}
\nonumber
\\
&=& 
\frac{\Gamma\left(-{D-1\over 2}\right)}{(4 \pi)^{(D-1)\over 2}}
\frac{l^{(z-1)(D-1)}}{L}
\left(\frac{2\pi}{L}\right)^{z(D-1)}
\frac{z(D-1)}{2\pi i} 
\textrm{tr}_{(\cal{R})}
\oint_C d w \, w^{z(D-1)-1} \ln [F(w,\alpha)],
\nonumber\\
\end{eqnarray}
where
\begin{eqnarray}
F(w,\alpha) &=& \cos(2\pi w) - \cos(2\pi \alpha),
\end{eqnarray}
and the contour $C$ is the set of the circles each of which surrounds 
$w = m \pm \alpha > 0$, $m=0,1,\cdots$, on the $w$-plane. 
Since the integrand is regular except for roots on the real axes,
we can change a closed path of $\textrm{Re}~w > 0$ to the contour
shown in Fig. \ref{contour-two},
where the closed path consists of the infinite semicircle on the  plane $C_1$,
the straight lines $C_2$ from $+i\infty$ to $+i\delta$ ($\delta$ is a positive infinitesimal) 
and $C_4$ from $-i\delta$ to $-i\infty$, and a semicircle $C_3$ centered  at the origin with radius $\delta$ .
The integrals along the infinite semicircle vanishes in the sense of analytic continuation 
and by taking the limit $\delta\to0$,
only integrals along $C_2$ and $C_4$ remain non-vanishing, and thus we obtain
\begin{eqnarray}
\lefteqn{
V_{\rm scalar}^{R^{D-1}\times S_z^1}(\alpha; M_S=0)} 
\nonumber\\
&=&
-\frac{\Gamma\left(-{D-1\over 2}\right)}{(4 \pi)^{(D-1)\over 2}}
\frac{l^{(z-1)(D-1)}}{L}
\left(\frac{2\pi}{L}\right)^{z(D-1)}
\frac{z(D-1)}{2\pi i} 
\nonumber\\&& \times
\tr_{({\cal R})}
\biggl[
- i^{z(D-1)}   \int_0^\infty d t \, t^{z(D-1)-1} \ln [F(it,\alpha)]
\nonumber\\&&
\phantom{MM} + (-i)^{z(D-1)}\int_0^\infty d t \, t^{z(D-1)-1} \ln [F(-it,\alpha)]
\biggr]
\nonumber\\
&=&
\frac{\Gamma\left(-{D-1\over 2}\right)}{(4 \pi)^{(D-1)\over 2}}
\frac{l^{(z-1)(D-1)}}{L}
\left(\frac{2\pi}{L}\right)^{z(D-1)}
\frac{z(D-1)}{\pi}
\nonumber\\&& \times 
\sin \left(\frac{\pi z (D-1)}{2} \right)
\tr_{({\cal R})}{\cal J},
\label{contour-streched}
\end{eqnarray}
where we have used $F(w,\alpha)=F(-w, \alpha)$ and 
\begin{equation}
{\cal J}\equiv
\int_0^\infty d t \, t^{z(D-1)-1}\ln [F(it,\alpha)].
\label{neweqone}
\end{equation}
Recalling a relation 
$\cosh(A)-\cos(B) = e^A (1 - e^{-A+iB})(1 - e^{-A-iB})/2$,
we can rewrite ${\cal J}$ as
\begin{eqnarray}
{\cal J}
&=&
\int_0^\infty d t \, t^{z(D-1)-1} 
\left\{ 
\ln (1-e^{-2\pi t + i 2\pi \alpha})+\ln(1-e^{-2\pi t - i 2\pi \alpha})
+ 2\pi t - \ln(2)
\right\}.
\end{eqnarray}
We note that the above integral contains divergences. We ignore such divergent terms because 
we are interested in $\alpha$-dependent ones. Using the expansion formula
\begin{eqnarray}
\ln(1 - e^{-2\pi t+i 2\pi n \alpha}) 
&=&-\sum_{n=1}^\infty \frac{e^{-2\pi nt+i 2\pi n \alpha}}{n},
\label{log-expand}
\end{eqnarray}
and after integration we get
\begin{eqnarray}
{\cal J} &=&
-\frac{2}{(2\pi)^{z(D-1)}} \Gamma(z(D-1))
\sum_{n=1}^\infty \frac{\cos(2\pi n \alpha)}{n^{z(D-1)+1}}.
\label{contour-result}
\end{eqnarray}
Combining \eqref{contour-streched} and \eqref{contour-result},
and  utilizing the Gamma function formulas, 
one can reproduce the same result as \eqref{V_s_05}.
%
%
%
\subsection{Use of Hypergeometric Functions}
%
%
%
Here we show the one more way to evaluate the effective potential.
Using Poisson summation, we can write (\ref{V_s_01}) as
%
\begin{eqnarray}
\lefteqn{
V_{\rm scalar}^{R^{D-1} \times S_z^1}(\alpha; M_S =0)} \nonumber\\
&=&
-\frac{\sqrt{2\pi} (2\pi)^{z(D-1)}}{(4\pi)^{\frac{D-1}{2}}}{1\over L}
\left( \frac{l^{z-1}}{L^{z}} \right)^{D-1}
\textrm{tr}_{(\cal{R})}\hspace*{-0.2cm}\sum_{n=-\infty}^\infty 
e^{i 2 \pi n \alpha} 
\int_0^\infty d \tau \, \tau^{-\frac{D-1}{2} - 1}
F_{2 z}\left(2\pi n ; \tau \right),
\label{effpot-integral}
\end{eqnarray}
%
where
%
\begin{equation}
F_{2 z}(y;\tau)\equiv
\frac{1}{\sqrt{2\pi}} 
\int_{-\infty}^\infty d x \, 
\exp[- \tau x^{2 z}] e^{i x y}
\end{equation}
%
and we have used the following Poisson summation formula
%
\begin{eqnarray}
\sum_{m=-\infty}^\infty \exp\left[-\tau (m+\alpha)^{2 z} \right]
&=&
\sqrt{2\pi}
\sum_{n =-\infty}^\infty
e^{i 2 \pi n \alpha}
F_{2 z} \left(2\pi n ; \tau \right).
\end{eqnarray}
We can rewrite $F_{2z}(y;\tau)$ into the form
%
\begin{align}
\label{F_{2z}_2}
F_{2z}(y;\tau)
 = \frac{1}{\sqrt{2\pi}z\tau^{1/(2z)}}
   \sum_{r=0}^{\infty}\frac{\displaystyle  \left(\frac{iy}{\tau^{1/(2z)}}\right)^{2r}}{(2r)!}
   \Gamma\Bigl(\frac{2r+1}{2z}\Bigr)\,,
\end{align}
%
where we have expanded $e^{ixy}$ in powers of $x$ and 
used the integral representation of the gamma function.
To proceed further, we note that any non-negative integer $r$
can uniquely be parameterized by two integers $s$ and $k$ such that
$r=zs+k$, where
$r = 0,1,2,\cdots$, $s = 0,1,2,\cdots$ and $k = 0,1,\cdots,z-1$.
In terms of the Pochhammer's symbol
%
\begin{align}
\label{PochhammerSymbol}
(a)_{s} 
 \equiv a(a+1)(a+2)\cdots(a+s-1) 
 =\frac{\Gamma(a+s)}{\Gamma(a)}\,,
\end{align}
%
we have
%
\begin{align}
\label{PochhammerSymbol_1}
\Gamma\Bigl(\frac{2r+1}{2z}\Bigr)
 &=\Gamma\Bigl(\frac{2zs+2k+1}{2z}\Bigr)
 = \Bigl(\frac{2k+1}{2z}\Bigr)_{\! s}\Gamma\Bigl(\frac{2k+1}{2z}\Bigr)\,,\\
\label{PochhammerSymbol_2}
(2r)!
 &=(2zs+2k)!
 = (2k)!\,\Bigl(\frac{2k+1}{2z}\Bigr)_{\! s}\Bigl(\frac{2k+2}{2z}\Bigr)_{\! s}
    \cdots\Bigl(\frac{2k+2z}{2z}\Bigr)_{\! s} \,(2z)^{2zs}\,.
\end{align}
%
It then follows that\footnote{
The authors would like to thank Professor H. So who taught us 
the derivation of Eq.~(\ref{F_{2z}_3}) given here.
}
%
\begin{align}
\label{F_{2z}_3}
F_{2z}(y;\tau)
 = &\frac{1}{\sqrt{2\pi}z\tau^{1/(2z)}}
   \sum_{k=0}^{z-1}\frac{\Gamma\bigl(\frac{2k+1}{2z}\bigr)}{(2k)!}
   \Bigl(\frac{iy}{\tau^{1/(2z)}}\Bigr)^{2k}\notag\\
   &\qquad\qquad\times
   {}_{1}F_{2z-1}\Bigl(1;\frac{2k+2}{2z},\frac{2k+3}{2z},\cdots,
   \frac{2k+2z}{2z};\Bigl(\frac{iy}{2z}\Bigr)^{2z}\frac{1}{\tau}\Bigr)\,,
\end{align}
%
where we have used the relation $(1)_{s}=s!$ and 
$_p F_q(\cdots)$ is the generalized hypergeometric function defined by
%
\begin{eqnarray}
{}_p F_q (a_1,\cdots,a_p; b_1,\cdots,b_q; w) 
&\equiv& 
\sum_{s=0}^\infty   
\frac{(a_1)_s (a_2)_s \cdots (a_p)_s}{(b_1)_s (b_2)_s \cdots (b_q)_s} \frac{w^s}{s!}\,.
\end{eqnarray}
%
Truncating the $n=0$ mode, which is independent to $\alpha$ and responsible for the UV divergence,
and integrating with respect to $\tau$, we obtain
%
\begin{eqnarray}
\lefteqn{
V_{\rm scalar}^{R^{D-1} \times S_z^1}(\alpha; M_S=0)
} \nonumber\\
&=&- \frac{2 \sqrt{2\pi} (2\pi)^{z(D-1)}}{(4\pi)^{\frac{D-1}{2}}}
\frac{1}{L}\left( \frac{l^{z-1}}{L^{z}} \right)^{D-1}
\textrm{tr}_{(\cal{R})}
\sum_{n=1}^\infty \cos(2 \pi n \alpha) 
G_{2 z}(2\pi n ,D),
\label{shiki77}
\end{eqnarray}
%
where $G_{2z}(y,D)$ is defined by
%
\begin{eqnarray}
G_{2z}(y,D)
&\equiv&
 \int_0^\infty d\tau \, \tau^{-\frac{D}{2}-\frac{1}{2}} F_{2z}(y;\tau).
\end{eqnarray}
%
By use of the formulas of an indefinite integral and an asymptotic expansion\footnote{
For $z \ge 2$, in the asymptotic expansions we have extra terms,
which are exponentially diverging and may spoil the finiteness of $G_{2z}(y,D)$ (and therefore the $\alpha$-dependent part of the effective potential may suffer from divergences).
However, by numerical studies we found that such divergences in $G_{2z}(y,D)$ are canceled out with each other in different $k$'s and finite values of $G_{2z}(y,D)$ are obtained for smaller $z$'s.
Therefore we have omitted such terms in \eqref{asymptotic_expansions}.
See Ref.~\cite{wolfram} for detailed asymptotic expansions of generalized 
hypergeometric functions.}
of generalized hypergeometric functions
%
\begin{align}
\label{indefinite_integral}
&\int du\, u^{\alpha-1}\,
       {}_{p}F_{q}(a_{1},\cdots,a_{p};b_{1},\cdots,b_{q};u)\notag\\
&\hspace{30mm}= \frac{u^{\alpha}}{\alpha}\,
       {}_{p+1}F_{q+1}(\alpha, a_{1},\cdots,a_{p};
       \alpha+1,b_{1},\cdots,b_{q};u),\\
\label{asymptotic_expansions}
&{}_{p}F_{q}(a_{1},\cdots,a_{p};b_{1},\cdots,b_{q};u)\notag\\
&\hspace{30mm}\stackrel{|u|\rightarrow \infty}{\longrightarrow}\ \ 
     \sum_{k=1}^{p}
     \frac{\Gamma(a_{k}) \left(\prod_{j=1}^{q}\Gamma(b_{j})\right)
           \prod_{j=1,j\ne k}^{p}\Gamma(a_{j}-a_{k})}
          {\left( \prod_{j=1}^{p}\Gamma(a_{j}) \right) 
           \prod_{j=1}^{q}\Gamma(b_{j}-a_{k})}
     (-u)^{-a_{k}},
\end{align}
%
we can show after some calculations that
%
\begin{align}
\label{G_{2z}}
G_{2z}(y,D)
 = \frac{2^{D-1/2}}{y^{z(D-1)+1}}\ 
   \frac{\Gamma(D/2)\Gamma(z(D-1)+1)}{\Gamma(D)}\ 
   \frac{\sin \bigl(\frac{z(D-1)}{2}\pi\bigr)}
        {\sin \bigl(\frac{(D-1)}{2}\pi\bigr)}.
\end{align}
%
Inserting it into Eq.~(\ref{shiki77}) leads to Eq.~(\ref{V_s_05}),
as expected.

\section{Effective Potential for Massive Matter}
%
%
In this appendix we show the calculation for \eqref{V_s_08} in detail.
Let us start with
\begin{eqnarray}
\lefteqn{V_{\rm scalar}^{R^{D-1}\times S_z^1}(\alpha,M)}
\nonumber\\
&=&  \int \frac{d^{D-1}p_{E}}{(2\pi)^{D-1}}\ 
    \frac{1}{L} \sum^{\infty}_{m=-\infty}
    \textrm{tr}_{(\cal{R})}\biggl[
    \ln\Bigl( p_{E}^{\ 2} + l^{2(z-1)}\Bigl(
    \frac{2\pi(m+\alpha)}{L}\Bigr)^{2z}+M^2\Bigr)\biggr]
    \nonumber\\
&=&
\frac{2(\sqrt{\pi})^{D-1}}{(2\pi)^{D-1}\Gamma(\frac{D-1}{2})} \frac{1}{L} 
\nonumber\\&& 
\times
\sum_{m=-\infty}^\infty\int_{M}^\infty d E \, E (\sqrt{E^2-M^2})^{D-3}
{\rm tr}_{({\cal R})} \ln \left[E^2 + l^{2z-2} \left(\frac{2\pi(m+\alpha)}{L}\right)^{2z}\right].
\end{eqnarray}
For odd $D$, we can write
\begin{equation}
V_{\rm scalar}^{R^{D-1}\times S_z^1}(\alpha, M)
=\frac{2(\sqrt{\pi})^{D-1}}{(2\pi)^{D-1}\Gamma(\frac{D-1}{2})} \frac{1}{L}
\sum_{q=0}^{\frac{D-3}{2}}
{}_{\frac{D-3}{2}} C_q \, (-M^2)^{\frac{D-3}{2} - q} \cdot I_q,
\label{V-Iq}
\end{equation}
where ${}_n C_k \equiv n!/(k!(n-k)!)$ is the binomial coefficient.
$I_q$ is given by
\begin{eqnarray}
I_q 
&\equiv&
\sum_{m=-\infty}^\infty\int_{M}^\infty d E \, E^{2q+1}
{\tr}_{({\cal R})} 
\ln \left[E^2 + l^{2z-2} \left(\frac{2\pi(m+\alpha)}{L}\right)^{2z}\right]
\nonumber
\\
&=& \left( \frac{l^{z-1}}{L^z}\right)^{2q+2} 
\int_{M L^z/l^{z-1}}^\infty dQ~~Q^{2q+1} \tr_{({\cal R})}W,
\end{eqnarray}
where
\begin{equation}
W \equiv
\sum_{m=-\infty}^\infty \ln \left[ Q^2 + (2\pi (m + \alpha))^{2z} \right].
\end{equation}
Here we have ignored $\alpha$-independent terms. 
Since the derivative $d W/d Q$ is given by
\begin{eqnarray}
\frac{d W}{d Q} 
&=&
\sum_{m=-\infty}^\infty \frac{2 Q}{Q^2 + (2\pi(m + \alpha))^{2z}}
= \sum_{j=1}^{2 z} \frac{G'_j(Q)}{G_j(Q)},
\label{dWdQ}
\\
G_j(Q) &\equiv& \sin \left( \pi \alpha - \frac{\omega_j Q^{1/z}}{2} \right),
\\
\omega_j &\equiv& \exp \left( \frac{i(2j-1)\pi}{2 z}\right),
\end{eqnarray}
(the derivation of \eqref{dWdQ} is given in the latter part of this appendix), we obtain
\begin{eqnarray}
W 
&=&
\ln \left[ \prod_{j=1}^{2z} G_j (Q) \right]
\nonumber\\
&=&
\ln\biggl[
\prod_{j=1}^{z} 
\sin\left(\pi\biggl(\alpha-\frac{Q^{1/z}\omega_j}{2\pi}\biggr)\right)
\sin\biggl(\pi\biggl(\alpha-\frac{Q^{1/z}\omega_j^*}{2\pi}\biggr)\biggr)
\biggr],
\nonumber\\
&=& \ln 
\biggl[
\frac{1}{4^z} \prod_{j=1}^{z} \exp(s_j Q^{1/z}) 
\left\{ 1 - \exp[- Q^{1/z} s_j + i(2\pi\alpha - Q^{1/z}c_j)] \right\}
\nonumber\\ && \quad \times
\left\{ 1 - \exp[- Q^{1/z} s_j - i(2\pi\alpha - Q^{1/z}c_j)] \right\}
\biggr],
\end{eqnarray}
where we have used the relation $\omega_j=\omega_{2z+1-j}^*$, and 
\begin{eqnarray}
s_j &\equiv& \Im \omega_j = \sin\left(\frac{i(2j-1)\pi}{2 z} \right),
\\
c_j &\equiv& \Re \omega_j = \cos\left(\frac{i(2j-1)\pi}{2 z} \right).
\end{eqnarray}
Neglecting some $\alpha$-independent parts in $W$ which correspond to the 
ultraviolet divergences, we write $I_q$ as 
\begin{eqnarray}
I_q 
&=& \tr_{({\cal R})}
\left( \frac{l^{z-1}}{L^z}\right)^{2q+2} \sum_{j=1}^{z} 2 \Re J_{q,j},
\label{Iq-Jq}
\\
J_{q,j} 
&\equiv& 
\int_{M L^z/l^{z-1}}^\infty d Q \, Q^{2q+1} 
\ln [1 - \exp(-(s_j + ic_j)Q^{1/z} + i2\pi\alpha)].
\end{eqnarray}
Using \eqref{log-expand} and
performing the integration of $J_{q,j}$, we get
\begin{eqnarray}
J_{q,j} 
&=&
- z(s_j + i c_j)^{-z(2q+2)} 
\sum_{n=1}^\infty 
\frac{e^{i 2\pi n \alpha}}{n^{z(2q+2)+1}}
\Gamma(z(2q+2), n\rho_j),
\label{Jq-result}
\end{eqnarray}
where $\Gamma(x,y) \equiv \int_{y}^\infty t^{x-1} e^{-x} d x$
is the incomplete Gamma function and 
$$
\rho_j=(s_j+ic_j)\left(\frac{ML^z}{l^{z-1}}\right)^{1/z}=\frac{i}{\omega_j}
\left(\frac{ML^z}{l^{z-1}}\right)^{1/z}.
$$
Combining \eqref{V-Iq}, \eqref{Iq-Jq} and \eqref{Jq-result},
we obtain
\begin{eqnarray}
\lefteqn{
V_{\rm scalar}^{R^{D-1}\times S_z^1}(\alpha,M)
} \nonumber\\
&=& - \frac{2(\sqrt{\pi})^{D-1}}{(2\pi)^{D-1} \Gamma(\frac{D-1}{2})} \frac{1}{L}
\sum_{q=0}^{\frac{D-3}{2}} {}_{\frac{D-3}{2}} C_{q} (-M^2)^{\frac{D-3}{2}-q}
\left( \frac{l^{z-1}}{L^z} \right)^{2q+2} 2z 
\nonumber\\&& \times
\tr_{({\cal R})}\sum_{j=1}^{z} \Re \Biggl\{
\left( -i\omega_j\right)^{z(2q+2)}
\sum_{n=1}^\infty \frac{e^{i 2\pi n\alpha}}{n^{z(2q+2)+1}}
\Gamma\left(z(2q+2), \frac{i n}{\omega_j} \frac{L M^{1/z}}{l^{(z-1)/z}}\right)
\Biggr\}.
\label{massive-1}
\end{eqnarray}
One can rewrite \eqref{massive-1} into more useful form.
Utilizing an expansion formula
$
\Gamma(n,x) = (n-1)! e^{-x} \sum_{k=0}^{n-1} x^k/k!
$,
we can rewrite $J_{q,j}$ as
\begin{eqnarray}
J_{q,j} &=& - z[z(2q+2)-1]! (-i\omega_j)^{z(2q+2)}
\sum_{n=1}^\infty \frac{e^{-n\rho_j + i2\pi n\alpha}}{n^{z(2q+2)+1}} 
\sum_{k=0}^{z(2q+2)-1} \frac{(n\rho_j)^k}{k!}
\nonumber
\\
&=&- z[z(2q+2)-1]! (-i\omega_j)^{z(2q+2)}
\sum_{k=0}^{z(2q+2)-1}
\frac{(\rho_j)^k}{k!} 
\Li_{z(2q+2)-k+1} (e^{-\rho_j + i2\pi\alpha}),
\label{Jq-2}
\nonumber\\
\end{eqnarray}
where we have changed the order of the summation and $\Li_s(x)$ is defined in (\ref{Li}).
Combining \eqref{V-Iq}, \eqref{Iq-Jq} and \eqref{Jq-2},
we obtain \eqref{V_s_08}.

\begin{figure}[htbp]
\centerline{\includegraphics[width=0.4\linewidth]{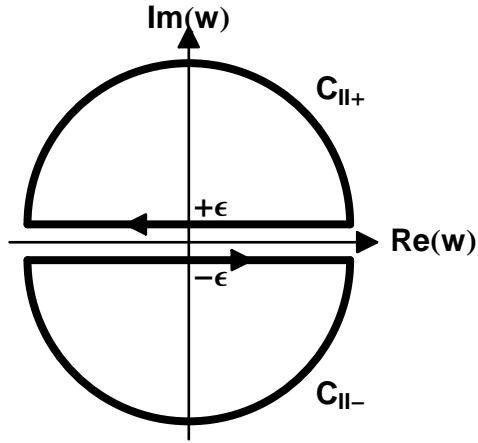}}
\caption{Two contours $C_{{\rm II}\pm}$. 
$C_{{\rm II}+}$ ($C_{{\rm II}-}$) consists of the infinite semicircle above (below) the line $\Im(w)=0$, and the 
straight path from $+\infty + i\epsilon$ ($-\infty - i\epsilon$)
to $-\infty +i\epsilon$ ($+\infty - i\epsilon$). 
$\epsilon$ denotes a positive infinitesimal.}
\label{B-cont2}
\end{figure}

Here we outline the derivation of \eqref{dWdQ}.
At first we point out that the left-hand-side of \eqref{dWdQ}
can be written in terms of the following contour integral:
\begin{eqnarray}
\sum_{m=-\infty}^\infty \frac{1}{x^{2z} + (2\pi (m+\alpha))^{2z}} 
 &=&
\frac{1}{2\pi i} \oint_{C_{\rm I}} d w \, \frac{\pi \cot(\pi w)}{x^{2z} + (2 \pi(w + \alpha))^{2z}} ,
\end{eqnarray}
where the path $C_{\rm I}$ denotes the set of the circles surrounding the points
$w = m$.
Without changing the value of the integral,
we can replace the integration path $C_{\rm I}$ by $C_{{\rm II}+}+C_{{\rm II}-}$ 
shown in Fig.~\ref{B-cont2}.
Counting the residues of the integrand at the points which are enclosed by $C_{{\rm II}\pm}$,
we obtain
\begin{eqnarray}
\frac{1}{2\pi i} \oint_{C_{\rm I}} d w \, \frac{\pi \cot(\pi w)}{x^{2z} + (2 \pi(w + \alpha))^{2z}} 
&=&
- \sum_{j=1}^{2z} \Res_{w \to \Omega_j}
\left[\frac{\pi \cot(\pi w)}{x^{2z} + (2\pi(w + \alpha))^{2z}}  \right],
\nonumber\\
&=& 
-\sum_{j=1}^{2z}\frac{\pi\cot\left(\pi\left(\frac{x\omega_j}{2\pi}-\alpha\right)\right)}
{2\pi x^{2z-1}\displaystyle{\prod_{k=1, k\neq j}^{2z}}(\omega_j-\omega_k)},
\label{sum-formula}
\end{eqnarray}
where $\Omega_j \equiv (x\omega_j-2\pi\alpha)/2\pi$,
and $\Res_{w\to a}[f(w)]$ denotes the residue of $f(w)$ at $w=a$.
One can reproduce \eqref{dWdQ} by plugging $x = Q^{1/z}$ to \eqref{sum-formula}
and using the formula
\begin{equation}
\displaystyle{\prod_{k=1,k\neq j}^{2z}}(\omega_j-\omega_k)=-\frac{2z}{\omega_j}.
\end{equation}
%
%
%
%

%
%
%
%
%
%
%
%
%
\end{document}